\documentclass[a4paper, final, 12pt]{article}

\usepackage{eurosym}
\usepackage{amsmath, amssymb, latexsym, amscd, amsthm,amsfonts,amstext}
\usepackage[mathscr]{eucal}
\usepackage{graphicx}
\usepackage{subfig}
\usepackage{float}
\usepackage{color}
\usepackage{hyperref}
\usepackage[utf8]{inputenc}
\usepackage[english]{babel}

\numberwithin{equation}{section}
\begin{document}

\title{Optimizing Stock Option Forecasting \\ 
with the Assembly of Machine Learning Models \\
and Improved Trading Strategies
}
\author{Zheng Cao$^1$, Raymond Guo$^2$, Wenyu Du$^3$, Jiayi Gao$^4$, \\
and Kirill V. Golubnichiy$^5$
\and University of Washington, Seattle, USA $^1$ $^2$ $^3$ $^4$
\and Department of Mathematics
\and zc68@uw.edu$^1$
\and Department of Computer Science \& Math 
\and rpg360@uw.edu$^2$
\and Department of Computer Science 
\and wenyudu@uw.edu$^3$
\and Academy for Young Scholars
\and jerrygao@uw.edu$^4$
\and University of Calgary, Alberta, Canada$^5$
\and Department of Mathematics and Statistics
\and kirill.golubnichi1@ucalgary.ca$^5$
}

\date{}
\maketitle

\begin{abstract}

This paper introduced key aspects of applying Machine Learning (ML) models, improved trading strategies, and the Quasi-Reversibility Method (QRM) to optimize stock option forecasting and trading results. It follows up on the findings from \textit{Application of Convolutional Neural Networks with Quasi-Reversibility Method Results for Option Forecasting}\cite{CNN}. First, the project included an application of Recurrent Neural Networks (RNN) and Long Short-Term Memory (LSTM) networks to provide a novel way of predicting stock option trends. Additionally, it examined the dependence of the ML models by evaluating the experimental method of combining multiple ML models to improve prediction results and decision-making. Lastly, two improved trading strategies and simulated investing results were presented. The Binomial Asset Pricing Model with discrete time stochastic process analysis and portfolio hedging was applied and suggested an optimized investment expectation. These results can be utilized in real-world trading strategies to optimize stock option investment results based on historical data.

\end{abstract}

\vspace{0.5em}


\newpage

\section{Introduction}

In their previous work, Cao, Du, and Golubnichiy \cite{CNN} developed a five-layer convolutional neural network that achieved $51.49\%$ accuracy and $57.14\% $ precision in testing. Here, the team expanded upon these results by using an LSTM for the same purposes and combining various neural networks to improve the stock option forecasting results. Updated trading strategies and simulated invested results were also illustrated for further analysis. The limitations and challenges were listed for future developments.

 To produce a more precise forecast of stock option pricing, Klibanov, Kuzhuget, and Golubnichiy created a new empirical mathematical modeling method \cite{KlibGol}. This could be accomplished using the Black-Scholes (BS) equation with initial and boundary conditions. The present value was calculated for a specific period of time in the past using the BS equation\cite{Hull}. For financial mathematics, it is a parabolic partial differential equation to predict the European style options that targets on the European style options\cite{ShreveB}. In this study, we employed the BS equation to forecast future option prices. The research thus belonged to the class of ill-posed problems, where the solution was either unstable or nonexistent. This research run into complications because we were attempting to forecast future option pricing. An ill-posed problem was one for which there was no solution or one that is unstable.

In order to apply regularization to solve the ill-posed problem, the system must be transformed into linear forms of functional equations.

In our case, $u$ is the minimizer, making it our prediction. We have a vector $X$, which contains 13 elements including the previous minimizer u and the the volatility coefficient ${\sigma}$ \cite[Chapter 7, Theorem 7.7]{Bjork}\textbf{:}

\begin{equation}
\begin{split}
& \frac{\partial u(s,\tau )}{\partial \tau }=\frac{{\sigma }^{2}}{2}s^{2}\frac{%
\partial ^{2}u(s,\tau )}{\partial s^{2}}, \\
& u(s,0)=f(s),
\end{split}
\label{1}
\end{equation}%
The payoff function is $f(s)=\max \left(
s-K,0\right) $, at T = t, where $K$ is the strike price \cite{Bjork}, $s>0.$, and the time at a given time $t$ will occur is $\tau$ 
\begin{equation}
\tau =T-t. 
\end{equation}
The option price function is defined by the Black-Scholes formula: 
\begin{equation}
u(s.\tau )=s\Phi (\theta _{+}(s,\tau ))-e^{-r\tau }K\Phi (\theta _{-}(s,\tau
)),
\end{equation}%

Based on the It\^o formula, we have:
\begin{equation}
du = (-\frac{\partial u(s, T-t)}{\partial \tau} + \frac{{\sigma}^2}{2} s^2 \frac{%
\partial^2 u(s, T-t)}{\partial s^2})dt + \sigma s \frac{\partial u(s, T-t)}{%
\partial s} dW.
\end{equation}
If equation (3) is solved forwards in time to forecast prices of stock options, it is an ill-posed inverse problem. By solving the previous equation, we get the solution as the minimizer and apply it into the trading strategy to generate prediction results. After testing on real market data, we proved all these new mathematical approaches suggest better prediction results which may help traders make better investment decisions.

The QRM, Binary Classification, and Regression Neural Network Machine Learning outcomes are summarized in the table below. The percentage of lucrative alternatives throughout model generations was shown in the Precision column of the table.

\bigskip \vspace{0.75cm} 
\textbf{Table 1. Previous Models' Results}
\begin{center}
\begin{tabular}{|l|l|l|l|}
\hline
Method & Accuracy & Precision & Recall \\ \hline
QRM & 49.77\% & 55.77\% & 52.43\% \\ \hline
Binary Classification & 56.36\% & 59.56\% & 70.22\% \\ \hline
Regression NN & 55.42\% & 60.32\% & 61.29\% \\ \hline
\end{tabular}
\end{center}

\section{RNNs and LSTMs}

\par
A Recurrent Neural Network (RNN) is a neural network that outputs both an preddiction (the attempted guess for a label) and a hidden state \cite{DeepLearning}. This hidden state is passed along to the RNN's next iteration, where it takes in both another input and the hidden state generated by the previous iteration. These hidden states allow RNNs to send themselves data containing information from previous predictions in order to determine the predictions that should be made in later iterations. In short, RNNs read through a series of tokens and have the benefit of being able to "remember past events" in order to make predictions.

This allows RNNs to be particularly effective when the input comes as a stream of data, where each datapoint in the stream has a label whose value is intuitively dependent on previous datapoints and labels. As a result, one of the main uses of RNNs is in predicting "time-series data," or sequences of datapoints that come in temporal order. Examples include predicting the weather, wins or losses of sports teams, or (for our applications) increases and decreases in stock prices.

In practice, traditional RNNs often have trouble using their hidden state to remember information over more than a few iterations. The Long Short-Term Memory network, or LSTM, was created to combat this issue. In every iteration, it passes itself two different hidden states instead of just one. One of these hidden states essentially undergoes the traditional processing of a hidden state in an RNN, and thus serves the role of the short-term memory. On the other hand, the second hidden state undergoes very few simple alterations in each iteration, and thus serves as the long-term memory. The creation of this second hidden state makes LSTMs significantly more likely to have the ability to remember information for more iterations. Since both short- and long-term information is necessary to make accurate predictions for option prices, we use LSTMs to attempt to make more accurate predictions.

Our architecture in this approach consisted of a two-layer LSTM followed by a fully connected layer and a sigmoid for classification. Our model was trained on sequences of 10 data points (10 consecutive days of stock information), fed in one at a time, with a batch size of 8.

This approach resulted in a $52.08\%$ accuracy in validation.

\section{ML Models Examination }

Based on previous modeling utilizing various machine learning techniques to help forecast stock options, this section explored experimenting with combining ML models for better results. The goal was to examine the dependence and inner correlation among the inner architectures of different ML models.

\subsection{Combing Prediction from Previous ML Models}
The motivation for improving the precision of the stock option predicting model was to combine the results of all previous models. Under the assumption that the models were executing independent decisions, the final results were expected to reach a higher degree of precision when all given models made the same decision.

Previous machine learning models had produced high precision and a high rate of profitable stock options, with the percentages for QRM, Binary Classification, RNN, and CNN models being 55.77, 59.56, 60.32, and 57.14 percent, respectively.

In addition, a base-case test was performed by determining the precision by combining the results from classification NN and model 10K CNN. With this combination, we were able to obtain a precision of 53.9 percent. Although we reached a precision that was lower than both original approaches, these results were biased due to the 13th factor, where both models generated results based on a common feature: the QRM model's results. This led to the models lacking independence from each other, which could explain the decrease in precision.

The table below lists one example of fusing Classification NN and CNN with the 10K model, which involved 10 thousand data rows of stock options.

\bigskip \vspace{0.75cm} 
\textbf{Table 1. Previous Models and Combined Model Results}
\begin{center}
\begin{tabular}{|l|l|}
\hline
Method & Precision \\ \hline
QRM & 55.77\% \\ \hline
Binary Classification & 59.56\% \\ \hline
Regression NN & 60.32\% \\ \hline
Classification (CNN) & 57.14\% \\ \hline
Classification NN + Model10k CNN & 53.9\% \\ \hline
\end{tabular}
\end{center}

\subsection{Experimental Analysis}

This approach was ultimately abandoned due to the fact that the ML models were dependent on each other. The output accuracy of 53.9\% performed worse than any of the models separately.

\begin{equation}
\begin{split}
& P_{1} := 0.56, \\
& P_{2} := 0.59, \\
& P := Joint Precision, \\
& P = \frac{0.56 * 0.59}{0.56 * 0.59 + (1 - 0.56)(1 - 0.59)} \\
& P \approx 0.647
\end{split}
\label{2}
\end{equation}%

A more general formula was shown as below:

\begin{equation}
\begin{split}
& P_{1} := Precision of Model 1, \\
& P_{2} := Precision of Model 2, \\
& P := Joint Precision, \\
& P = \frac{P_{1} * P_{2}}{P_{1} * P_{2} + (1-P_{1}) * (1-P_{2})} \\
\end{split}
\label{3}
\end{equation}%

This confirmed the hypothesis that there existed nontrivial correlation among the ML models and dependence of the prediction results. Given two 100\% independent forecasting results, with 0.56 and 0.59 precision, the expected combined model should return a value of 0.647.

\section{Improved Trading Model}

To utilize the ML forecasting results into real-world application, we must develop trading strategies to help execute investing decisions and simulate expected returns.

\subsection{Previous Trading Strategy}

Cao, Du, and Golubnichiy introduced a simple and straight-forward trading method in their previous paper.\cite{CNN}Let’s denote $s$ as the stock price, $t$ as the time, and $\sigma(t)$ as the volatility of the option. We are going to be using the historically implied volatility based on the market data of [2]. here, we assume that $\sigma = \sigma(t)$ in order to avoid other historical data for the volatility. Let’s call $ub(t)$ and $ua(t)$ the bid and ask prices of the options at the moment of time $t$ and $sb(t)$ and $sa(t)$ the bid and ask prices of the stock at the moment of time $t$. We will be buying in an option given that the following holds:
$EST(\tau) \ge REAL(0)$.

\subsection{Trading Strategy Advancements}

This paper presented new trading strategy advancements in addition to the previous efforts, including the investing methods and simulated results.

\subsubsection{Trading Strategy Simulation}

This first segment initiated trading decisions based solely on machine learning prediction results. Simulating trading applied the previous trading strategy and the precision of our estimators, and we saw profits shown in the following graphs:

\begin{center}

\includegraphics[scale=0.7]{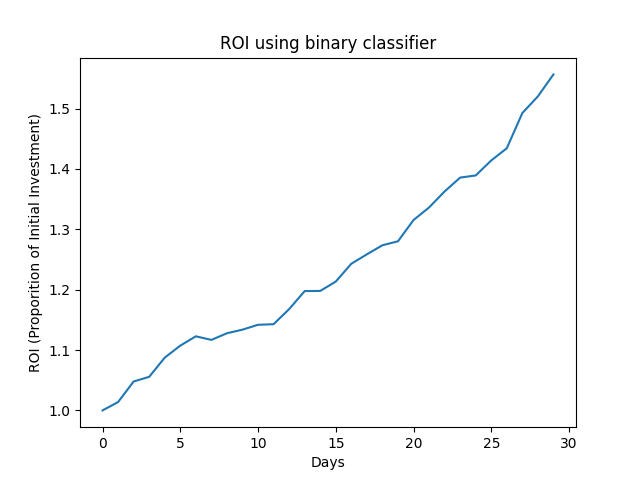}

\includegraphics[scale=0.7]{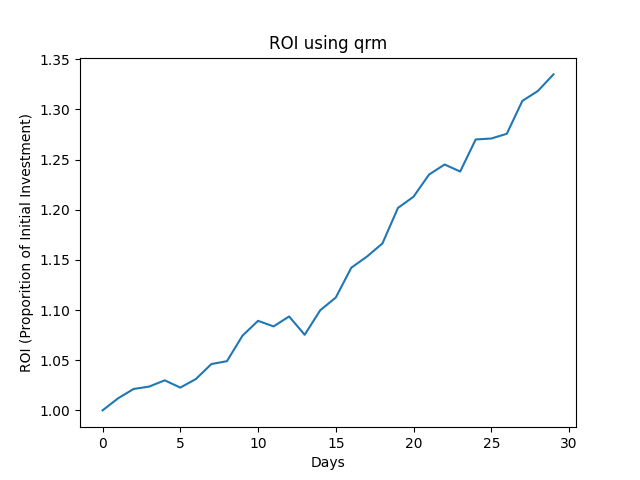}

\includegraphics[scale=0.7]{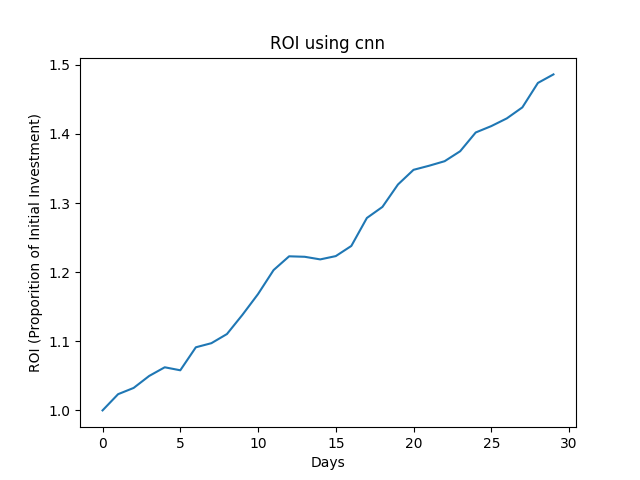}

\end{center}

\subsubsection{Binomial Trading Model}
This trading strategy is an improved version compared to the previous developed by Dr. Golubnichiy and his colleagues. It was inspired from the Binomial Asset Pricing Model as introduced by Dr Shreve in his book, "Stochastic Calculus for Finance I: The Binomial Asset Pricing Model" \cite{ShreveA}. The goal of this modeling was to compute the expectation of how a given portfolio of stock options would perform at a certain day, when executing the ML predicting results. 

\par
To begin with, let's define some variables as the followings:
\begin{itemize}
    \item ROR: Rate of return
    \item ROL: Rate of loss
    \item p: precision, in $\%$
    \item $M_k$: The expected asset we get on day k.
    \item $M_k, _x$: The expected asset for the $x^th$ scenario on day k.
\end{itemize}
\par
For rate determination, we determined the rates based on the probability and the magnitude to which the portfolio went up. For the convenience of programming, we will be setting a fixed probability P, defined by the model precision.

\includegraphics[scale=0.8]{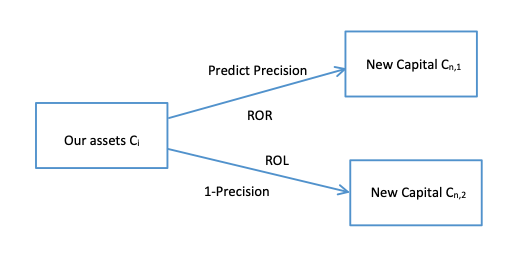}
\par
This analysis begins with the assumption that we have infinite money, and uses $C_i$ as the variable for initial assets. Note that here we utilized the current CNN precision percentage, $56\%$ precision, as an example to help demonstrate the process.
\par
After each trading day, each possibility of our assets would turn out to have two new possible outcomes, $C_n, _1$ and $C_n, _2$.
\par
These two values could be reached depending on the precision and (precision minus 1) of our ML models. Although bias and error rates for prediction accuracy across each company might be different, with enough data, these minor variances could be balanced out. Furthermore, we could treat all of the companies as a single entity, as a portfolio, with a 56\% chance of increasing (from CNN), for example.
\par
Therefore, with $56\%$ chance that our prediction to be accurate, we would have a $44\%$ chance that when we invested into a company or portfolio we predicted would be profitable, it would instead lead to a loss.
\par
\includegraphics[scale=0.8]{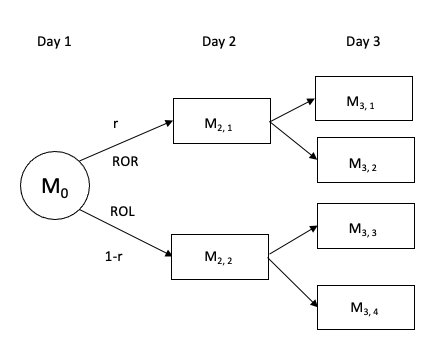}
\newline
As illustrated from the above image, are an increasing number of outcomes possible after each day, with our initial capital $M_0$ and then on the $k_th$ day, we would have a capital of $M_k$. This value was calculated by multiplying the probability $p$ or $1-p$ along the path from the initial day by the expected capital in $m_k, _x$. Here, our $m_k, _x$ was obtained by multiplying the previously connected index by the ROR.
\par
The ROR, or Rate of Return, would be calculated through the Black-Scholes model, which predicted the next day's prices. To get ROR, we summed up all of the worth of the capital as predicted by the Black-Scholes the next day and divided by the total capital for today.
\par
For ROL, it would hence be natural to claim that it would equal to ROR. 


\par
\includegraphics[scale=0.7]{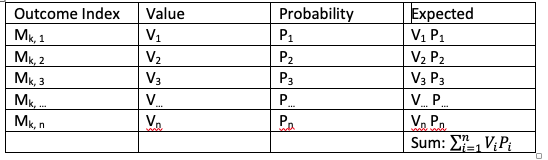}
\newline
After the number of trading days, our final result would have a huge list of different possible capitals and their respective probabilities throughout all the days, such as the one above. We then multiplied the probability by the capital and sum up all of these values. This return the final expected value of return after a set number of trading days. 
\par
As the example illustrated from the above picture, in just one trading day, the expected value of return would be $C_n,_1 * 56\% + C_n, _2 * 44\%$.
\par
As a better view of how the steps work throughout the entire day, suppose we had the predicted precision P of 2/3, the ROR of 2, and the ROL of 0.5. This was possible given that, when we were merging different combinations, the BS model would output different values. The entire binomial model was a martingale thus its expectation could be computed by applying Wald's Theorem.

\section{Limitations and Future Developments}

This section addresses several challenges that occurred during the research.

First, the LSTM was limited in its use case since it was trained solely on sequences of fixed length (10). This meant the LSTM might perform unpredictably if it was programmed to take in significantly longer sequences of tokens.

Additionally, the independent models might be optimized by manipulating the data so that the models will be independently trained. When independent ML models are used and combined to predict future trends of stocks and options, we can expect an increase in precision.

Also, the trading strategies had great potential for development. For each node in the binomial trading tree, we did not apply advanced techniques such as the Hamilton–Jacobi–Bellman (HJB) equation to achieve asset hedging based on their historical data. Instead, we chose to apply a relatively simple and straightforward method, as introduced earlier. Even though this model returned a comparatively lower approximation accuracy, it indeed functioned as a more efficient algorithm for users to calculate expected profit returns based on the previous models introduced in each section. For future developments, we would consider applying various hedging models and statistical equations to program a more accurate model, as suggested by Professor Zhen-Qing Chen.

\section{Acknowledgement}

We are very grateful to Professors Zhen-Qing Chen and Fang Han from the University of Washington for providing this study with various modeling techniques and strategy building inspiration and direction. We are also delighted for the development of QRM and other modeling techniques by Professor Michael V. Klibanov from University of North Carolina at Charlotte, Charlotte.

\section{Summary}

In conclusion, the LSTM outperformed QRM in terms of stock price prediction, however its precision lagged below some earlier models.  Multiple ML techniques were demonstrated to be ineffective when combined, suggesting that there existed nontrivial relationships between various models.The findings of ML simulated trading and the Binomial Asset Pricing Model were presented, demonstrating the advances made during this research stage and the viability of the investing strategies in the actual U.S. stock option market.

\newpage

%
%

\end{document}